\documentclass{elsart}

\usepackage{graphicx}
\usepackage{subfigure}
\usepackage{amssymb}
\usepackage[dvips]{color}

\begin{document}

\begin{frontmatter}

\title{Extended moment formation and magnetic ordering in the trigonal 
       chain compound $ {\rm \bf Ca_3Co_2O_6} $}

\author{V.\ Eyert$^1$\corauthref{corauth}}, 
\corauth[corauth]{Corresponding author. fax: +49 821 598 3262} 
\ead{eyert@physik.uni-augsburg.de}
\author{C.\ Laschinger$^1$}, 
\author{T.\ Kopp$^1$}, 
\author{R.\ Fr\'{e}sard$^2$}

\address{$^1$Institut f\"ur Physik, Universit\"at Augsburg, 
             86135 Augsburg, Germany, \\
         $^2$Laboratoire Crismat, UMR CNRS-ENSICAEN (ISMRA) 6508, 
             Caen, France}

\begin{abstract}
The results of electronic structure calculations for the one-dimensional
magnetic chain compound $ {\rm Ca_3Co_2O_6} $ are presented. The calculations
are based on density functional theory and the local density approximation
and used the augmented spherical wave (ASW) method. Our results allow for
deeper understanding of recent experimental findings. In particular,
alternation of Co $ 3d $ low- and high-spin states along the characteristic
chains is related to differences in the oxygen coordination at the inequivalent
cobalt sites. Strong hybridization of the $ d $ states with the O $ 2p $
states lays ground for polarization of the latter and the formation of
extended localized magnetic moments centered at the high-spin sites. In
contrast, strong metal-metal overlap along the chains gives rise to
intrachain ferromagnetic exchange coupling of the extended moments via 
the $ d_{3z^2-r^2} $ orbitals of the low-spin cobalt atoms.
\end{abstract}

\end{frontmatter}

\section{Introduction}

Low-dimensional physical systems are attracting a lot of attention since long 
due to the occurence of exciting physical properties deviating distinctly 
from those known from three-dimensional systems. This includes both, new 
types of ordering phenomena being due to, e.g., Fermi-surface instabilities 
and new types of excitations as resulting, e.g., from spin-charge separation. 
Furthermore, being embedded in three-dimensional crystals, low-dimensional 
atomic arrangements give rise to different length scales and coupling 
strengths and thus may lead to rich phase diagrams. 
Finally, the existence of magnetic moments in such systems raises questions 
concerning the relevant exchange mechanisms and their anisotropies.  

In recent years, interest especially in one-dimensional systems has focused 
on a new class of transition metal oxides of the general formula 
$ {\rm A_3'ABO_6} $ ($ {\rm A' = Ca, Sr, Ba} $; $ {\rm A, B =} $ 
transition metal), which crystallize in the trigonal $ {\rm K_4CdCl_6} $ 
structure \cite{stitzer01}. The space group is $ R\bar{3}c $ ($ D_{3d}^{6} $, 
No.\ 167). In these compounds, transition metal-oxygen polyhedra form 
well separated chains running parallel to the trigonal axis; see, e.g., 
Refs.\ \cite{fjellvag96,aasland97} for a representation of the crystal 
structure. Space between 
the chains is filled with the $ {\rm A'} $ cations. Each chain consists of 
alternating, face-sharing $ {\rm AO_6} $ trigonal prisms and $ {\rm BO_6} $ 
octahedra. Typically, the ratio of interchain to intrachain metal-metal 
distance is of the order of two, this fact explaining the  pronounced 
one-dimensionality. The chains themselves are arranged on a triangular 
lattice. As a consequence, in addition to showing the abovementioned unique 
properties these compounds allow to study geometric frustration effects 
as partial disorder and spin-glass like behaviour. 

Prominent members of this class are the cobaltates $ {\rm Ca_3CoBO_6} $ 
with $ {\rm B=Co} $, $ {\rm Rh} $, $ {\rm Ir} $, which have gained much 
interest due to their exciting magnetic properties. Here we concentrate 
on $ {\rm Ca_3Co_2O_6} $, which has two inequivalent cobalt sites, one 
in octahedral environment 
(labelled Co1) and the other (Co2) centered in the trigonal prisms 
\cite{fjellvag96,aasland97}. According to powder neutron diffraction data  
reported by Fjellv\aa g {\em et al.}\ room-temperature lattice constants 
amount to $ a_{hex} = 9.0793 $\,\AA \ and $ c_{hex} = 10.381 $\,\AA 
\cite{fjellvag96}. Data taken at 10\,K by Aasland {\em et al.}\ resulted in 
$ a_{hex} = 9.060 $\,\AA \ and $ c_{hex} = 10.366 $\,\AA \ with only slight 
changes of the atomic positional parameters as compared to 298\,K 
\cite{aasland97}. The latter data were used in the calculations. 

While the Co--Co distance within the chains amounts to 2.595\,\AA, 
next-nearest neighbour distances across the chains are 5.31\,\AA \ 
\cite{fjellvag96}. This value is somewhat larger than the interchain 
separation of 5.24\,\AA, since neighbouring chains are shifted parallel 
to the trigonal axis by 1/3 of the intrachain separation. According 
to the neutron diffraction data, 
the octahedral Co1-O distances of 1.916\,\AA \ are considerably shorter 
than the Co2-O distances, which amount to 2.062\,\AA \cite{fjellvag96}. 
From the difference it was expected that $ {\rm Ca_3Co_2O_6} $ might 
show charge or spin order at low temperatures \cite{fjellvag96}. Finally, 
the short Co1--Co2 distance within the chains, which is a result of the 
face-sharing of the polyhedra and is close to the value of 2.51\,\AA \ 
in metallic cobalt, has been taken as indicative for strong metal-metal 
bonding \cite{fjellvag96}. 

Magnetic susceptibility data revealed the onset of magnetic ordering below 
$ {\rm T_{C1}} = 24 $\,K \cite{aasland97,kageyama97a,maignan00}. In 
addition, a second transition was found at $ {\rm T_{C2}} = 12 $\,K 
\cite{kageyama97a,maignan00}. From Curie-Weiss behaviour above 80\,K 
an effective magnetic moment of $ 5.7 \mu_B $ and a paramagnetic Curie 
temperature of 28\,K pointing to predominant ferromagnetic 
ordering was inferred. Neutron diffraction measurements by Aasland 
{\em et al.}\ revealed local magnetic moments of $ 0.08 \mu_B $ and 
$ 3.00 \mu_B $ for octahedral Co1 and trigonal 
prismatic Co2, respectively \cite{aasland97,kageyama97a}. All moments are 
aligned along the trigonal axis. From these data it was furthermore 
concluded that the coupling is ferromagnetic within the chains and 
antiferromagnetic across the chains \cite{aasland97}. While 
$ {\rm T_{C1}} $ has been associated with the ferromagnetic intrachain 
coupling, $ {\rm T_{C2}} $ would correspond to long-range three-dimensional 
ordering \cite{kageyama97a,maignan00}. 

Magnetization measurements showed a plateau at $ \approx 1.31 \mu_B $ per 
f.u.\ for low field and a steep increase by a factor of three at about 
3.5\,T, which was interpreted as a ferri- to ferromagnetic transition 
\cite{aasland97,kageyama97a}. However, the final magnetic moment of about 
$ 4 \mu_B $ was difficult to reconcile with the values found for the 
local cobalt moments \cite{aasland97,kageyama97c}. While Aasland 
{\em et al.}\ proposed a magnetic structure with 1/3 of the chains being 
antiparallel to the remaining 2/3 \cite{aasland97}, Kageyama {\em et al.}\ 
argued that at low field and temperatures between 10 and 24\,K the system 
assumes a partially disordered antiferromagnetic (PDA) state 
\cite{kageyama97a}. The existence of this state in different materials is 
still a matter of controversal discussion. While it was also reported for 
$ {\rm Ca_3CoRhO_6} $ \cite{niitaka01}, recent magnetization and heat 
capacity data allowed to exclude the PDA state for $ {\rm Ca_3CoIrO_6} $ 
\cite{rayaprol03}. For $ {\rm Ca_3Co_2O_6} $ it was shown that pressure 
leads to increase of $ {\rm T_{C2}} $ and, hence, stabilization of the 
ferrimagnetic phase \cite{martinez01,hernando02}. Magnetization data taken 
below $ {\rm T_{C2}} $ on single 
crystals by Maignan {\em et al.}\ displayed a sequence of up to four 
different magnetization plateaus pointing to a variety of competing 
magnetic structures \cite{maignan00}. Moreover, 
AC-$ \chi $ measurements by Maignan {\em et al.} showed a remarkably 
strong frequency shift, which has been interpreted as a measure of the 
frustration between different magnetic configurations \cite{maignan00}. 
These results supported interpretations of $ {\rm Ca_3Co_2O_6} $ as a 
triangular lattice of antiferromagnetically coupled Ising spins, each 
formed from a single chain. However, the exact nature of the magnetic 
ground state of this material is still unclear. 

In this letter we report on electronic structure calculations for 
$ {\rm Ca_3Co_2O_6} $, which include the spin-degenerate case as well as 
spin-polarized configurations assuming both the spin-structure proposed 
by Aasland {\em et al.}\ and the hypothetical case of ferromagnetic 
alignment of the chains. Our calculations reveal i) low- and high-spin 
moments, respectively, at the octahedral and trigonal prismatic cobalt 
sites, ii) a rather large oxygen moment due to polarization by the high-spin  
cobalt sites, which together give rise to the formation of extended but 
still well localized $ {\rm CoO_6} $ moments and resolve the discrepancy 
between the magnetization and neutron diffraction data, and iii) coupling 
of these extended moments by ferromagnetic exchange via the $ 3d $ states 
of the low-spin octahedral cobalt atoms.

\section{Methodology}

The calculations were performed using the scalar-relativistic augmented  
spherical wave (ASW) method \cite{wkg,revasw}. In order to represent the 
correct shape of the crystal potential in the large voids of the open 
crystal structure, additional augmentation spheres were inserted. Optimal 
augmentation sphere positions as well as radii of all spheres were
automatically generated by the sphere geometry optimization (SGO) 
algorithm \cite{eyert98b}.
Self-consistency was achieved by an efficient algorithm for convergence
acceleration \cite{mixpap}. Brillouin zone sampling was done using an
increased number of $ {\bf k} $-points ranging from 28 to 408 points 
within the irreducible wedge.

\section{Results and Discussion}

\subsection{Spin-Degenerate Calculations}

\begin{figure}[htb]
\centering 
\subfigure{\includegraphics[width=0.48\textwidth]{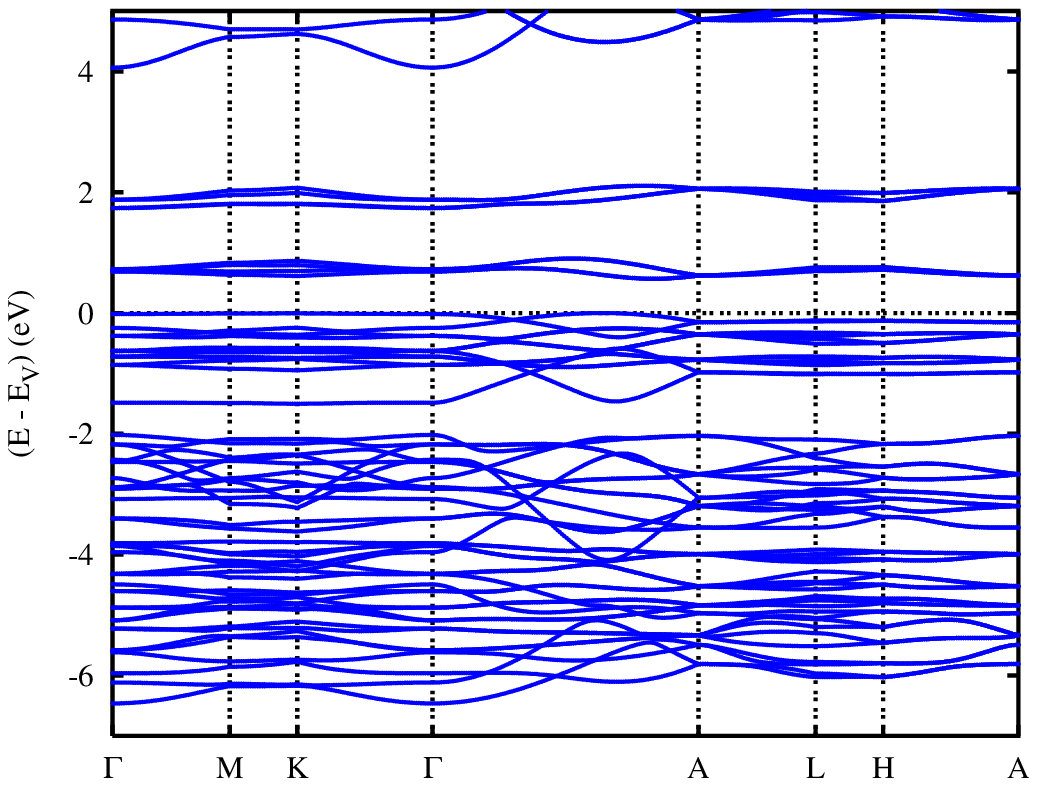}}
\subfigure{\includegraphics[width=0.48\textwidth]{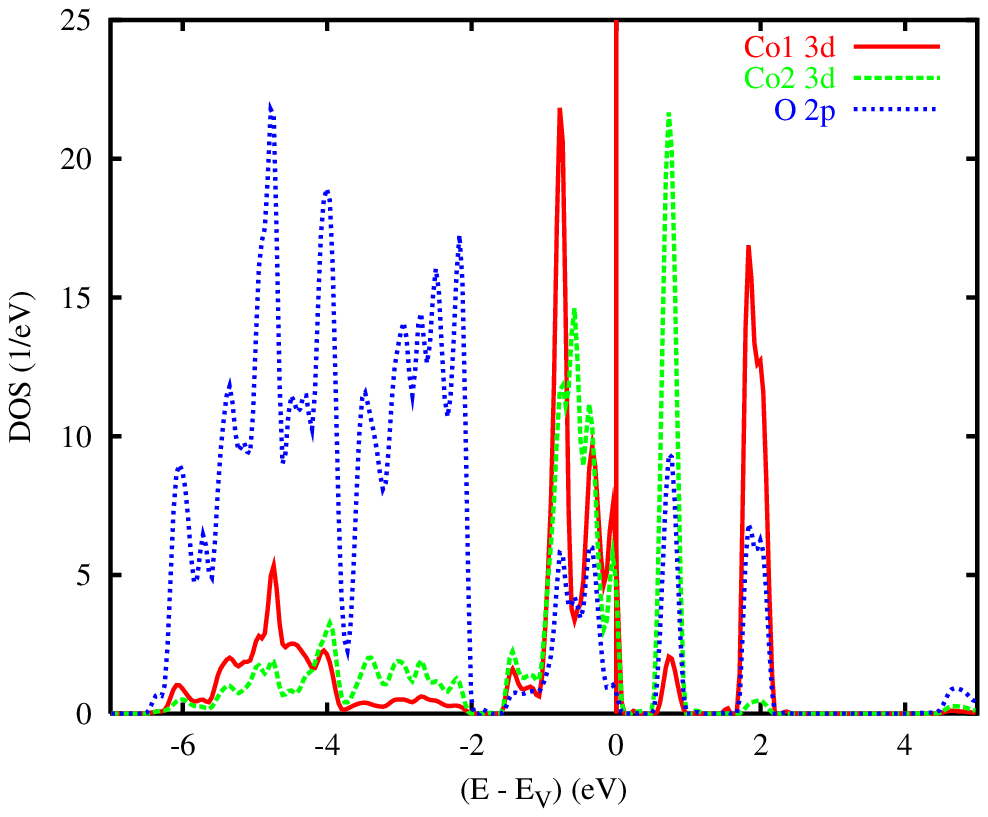}}
\caption{Electronic bands and partial DOS of nonmagnetic 
         $ {\rm Ca_3Co_2O_6} $.}
\label{fig:res1}
\end{figure}

In a first step the electronic properties of $ {\rm Ca_3Co_2O_6} $ were 
calculated with spin-degeneracy enforced. The resulting electronic 
structure and partial densities of states (DOS) are displayed in Fig.\ 
\ref{fig:res1}. 
All bands display rather strong dispersion parallel to the chain axis, 
i.e.\ along the line $ {\rm \Gamma} $-A, reflecting the  
one-dimensionality of the compound. However, the finite perpendicular 
dispersion points to albeit weak interchain coupling. 

Five groups of bands are identified. In the energy range from -6.4 to 
-2.0\,eV 36 bands are observed, which trace back mainly to the O $ 2p $ 
states. In contrast, the next three groups of bands, which extend from 
-1.7\,eV to the valence band maximum, from 0.6 to 1.0\,eV, and from 
1.8 to 2.2\,eV, comprise twelve and two times four bands, respectively, 
and derive mainly from the Co $ 3d $ states. Finally, bands starting 
above 4\,eV are of Co $ 4s $ character. As is obvious from the partial 
DOS, $ p $--$ d $ hybridization causes substantial $ d $ and $ p $ 
contributions, respectively, below and above -2\,eV reaching about 40\%  
especially in the upper valence band and the conduction bands. Whereas 
$ d $ contributions from the octahedrally coordinated Co1 show up mainly 
below -3.8\,eV and give rise to the peak at -4.8\,eV, the Co2 $ d $ 
contributions are equally spread over the whole energy range between 
-6 and -2\,eV. In passing we mention the $ d $ admixture to the peaks 
at 0.8 and 2.0\,eV, which is of almost pure Co2 and Co1 character, 
respectively. From the partial DOS the cobalt atoms can be formally 
assigned a $ d^6 $ configuration, which, together with the absence of 
charge disproportionation, is in perfect agreement with the neutron 
scattering results \cite{aasland97}. 

\begin{figure}
\centering 
\subfigure{\includegraphics[width=0.48\textwidth]{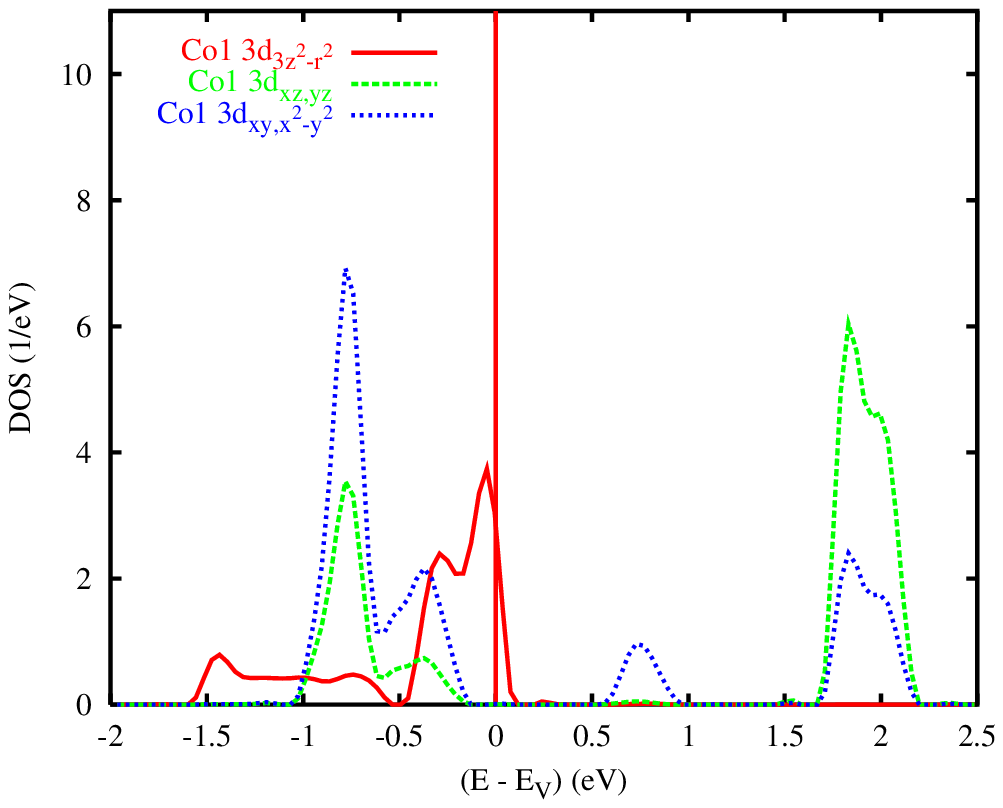}}
\subfigure{\includegraphics[width=0.48\textwidth]{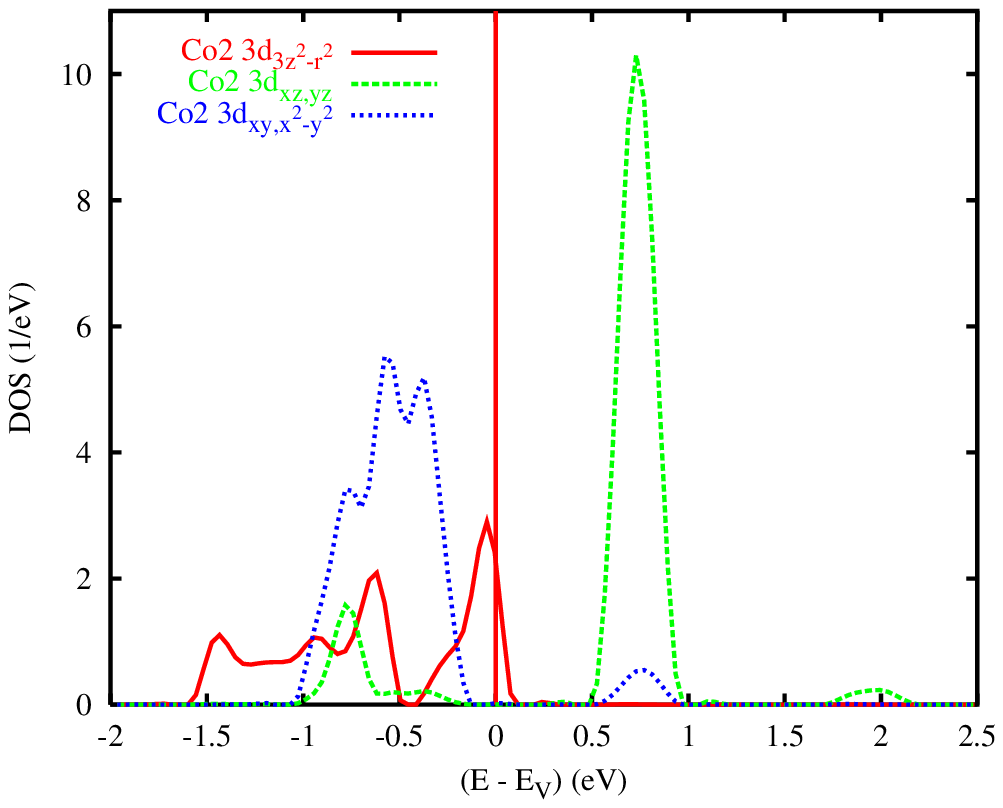}}
\caption{Partial Co1 and Co2 $ 3d $ DOS of nonmagnetic 
         $ {\rm Ca_3Co_2O_6} $.
         Slight broadening is due to the DOS calculation scheme
         \protect \cite{methfessel89}.}
\label{fig:res2}
\end{figure}

Crystal field splitting due to the octahedral and trigonal prismatic 
environment of Co1 and Co2 atoms, respectively, is revealed on closer 
inspection of the partial $ d $ DOS as displayed in Fig.\ \ref{fig:res2}. 
Note that we used for both sites the same global coordinate system with 
the $ z $ axis parallel to the chain axis. It deviates from the natural 
coordinate system for octahedral coordination with the Cartesian axes 
parallel to the metal-oxygen bonds due to rotation of the octahedra 
centered at the Co1 sites. As a consequence, both the $ t_{2g} $ and 
$ e_g $ bands arise as admixtures of all five $ d $ states (referred 
to the global coordinate system), a situation well known from 
other trigonal and hexagonal systems as, e.g., $ {\rm V_2O_3} $ 
\cite{held01}. Nevertheless, at the Co1 sites, the octahedral crystal 
field leads to nearly perfect splitting into fully occupied $ t_{2g} $ 
and empty $ e_g $ states. In contrast, the trigonal crystal field at 
the Co2 sites leads to splitting into non-degenerate $ d_{3z^2-r^2} $ 
as well as doubly degenerate $ d_{xy,x^2-y^2} $ and $ d_{xz,yz} $ states. 
While the latter appear almost exclusively in the peak around 0.75\,eV, 
the former three orbitals dominate the upper valence band. 

Apart from the abovementioned strong $ p $-$ d $ hybridization two 
findings are important for the understanding of the magnetic properties. 
First, as has already been mentioned, the lower and upper conduction band, 
respectively, centered at 0.8 and 2\,eV are due exclusively to the Co2 
$ d_{xz,yz} $ and Co1 $ e_g $ states. This is related to the fact that the 
latter states experience $ \sigma $-type $ p $-$ d $ bonding and, being 
antibonding, are thus pushed to higher energies 
whereas the $ p $-$ d $-bonding of the $ d_{xz,yz} $ orbitals is less strong. 
In the present situation this difference lets us expect that in a 
spin-polarized calculation the Co2 sites will carry larger magnetic moments. 
Second, we point to the large width of the $ d_{3z^2-r^2} $ bands as obvious 
from both the Co1 and Co2 partial DOS, which is indicative of the strong 
metal-metal overlap along the chain axis. This has been confirmed by the 
calculated crystal orbital overlap population (COOP) and covalent 
bond energies as well as a more detailed analysis of 
the electronic wave functions. According to the latter the $ d_{3z^2-r^2} $ 
states give rise to the dispersionless bands along  
$ {\rm \Gamma} $-M-K-$ {\rm \Gamma} $ at -1.5\,eV and at $ {\rm E_F} $ and 
show finite dispersion only parallel to $ {\rm \Gamma} $-A. Within these 
bands the bonding states are more Co2-like, whereas the antibonding states 
have increased Co1-like character.

\subsection{Spin-Polarized Calculations assuming Ferromagnetic Order}

In a second step, spin-polarized calculations were performed. Yet, in 
view of the predominant influence of the ferromagnetic interchain 
coupling as deduced from the experimental data we started out from the 
hypothetical situation, where all chains are aligned in parallel, 
before turning to the ferrimagnetic spin-structure proposed by Aasland 
{\em et al.}\ \cite{aasland97}. Insofar our procedure is along the same 
line of reasoning as the calculations by Whangbo {\em et al.}, who, 
however, did not deal with the observed ferrimagnetic structure and 
concentrated mainly on the origin of the cobalt moments \cite{whangbo03}. 

For the assumed ferromagnetic order we find a converged solution with 
well localized magnetic moments of 2.73, 0.35, and 0.14 $ \mu_B $, 
respectively, on Co2, Co1, and oxygen. These values reflect the 
experimental finding of a high and low spin state of the cobalt atoms 
at the trigonal prismatic and octahedral sites, respectively, and are 
close to those reported by Whangbo {\em et al.}\ as well as the neutron 
diffraction results of $ 3.00 $ and $ 0.08 \mu_B $ for the cobalt atoms. 
The total moment per unit cell, i.e.\ per two formula units, amounts to 
exactly $ 8.0 \mu_B $. Worth mentioning are the rather high magnetic moments 
associated with the oxygen sites. Summing up to almost 1\,$ \mu_B $ per 
formula unit, they provide an easy explanation for the discrepancy between 
the neutron scattering data and the magnetization in high field. 

\begin{figure}
\centering 
\subfigure{\includegraphics[width=0.48\textwidth]{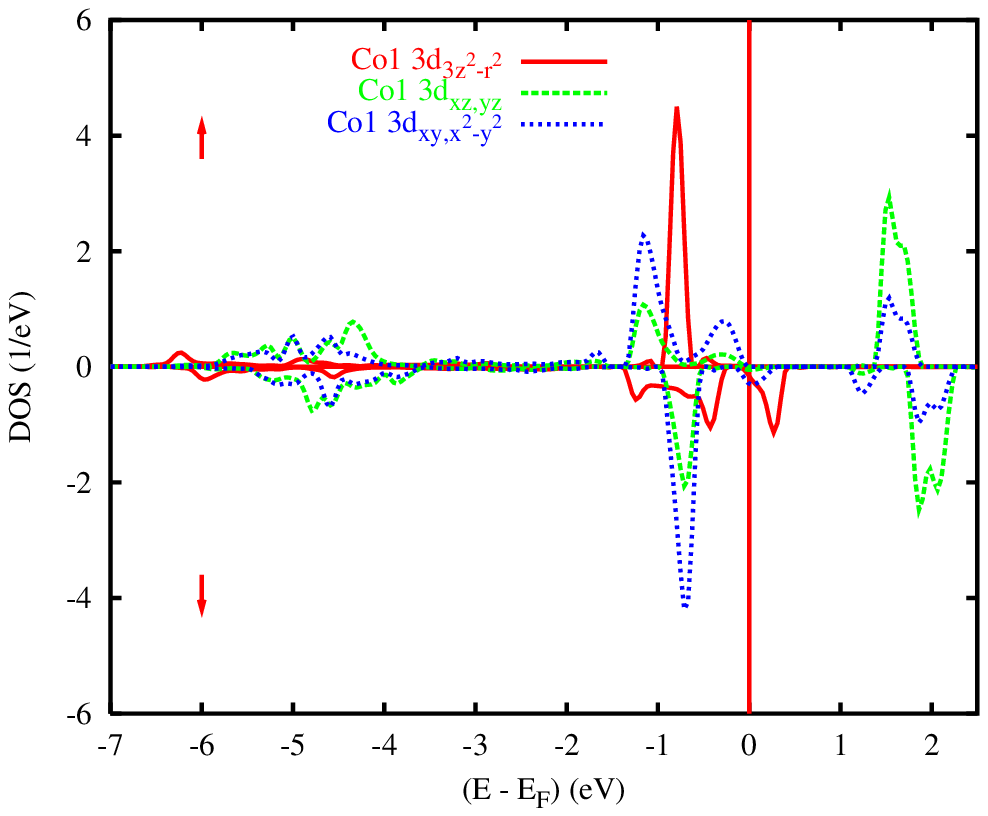}}
\subfigure{\includegraphics[width=0.48\textwidth]{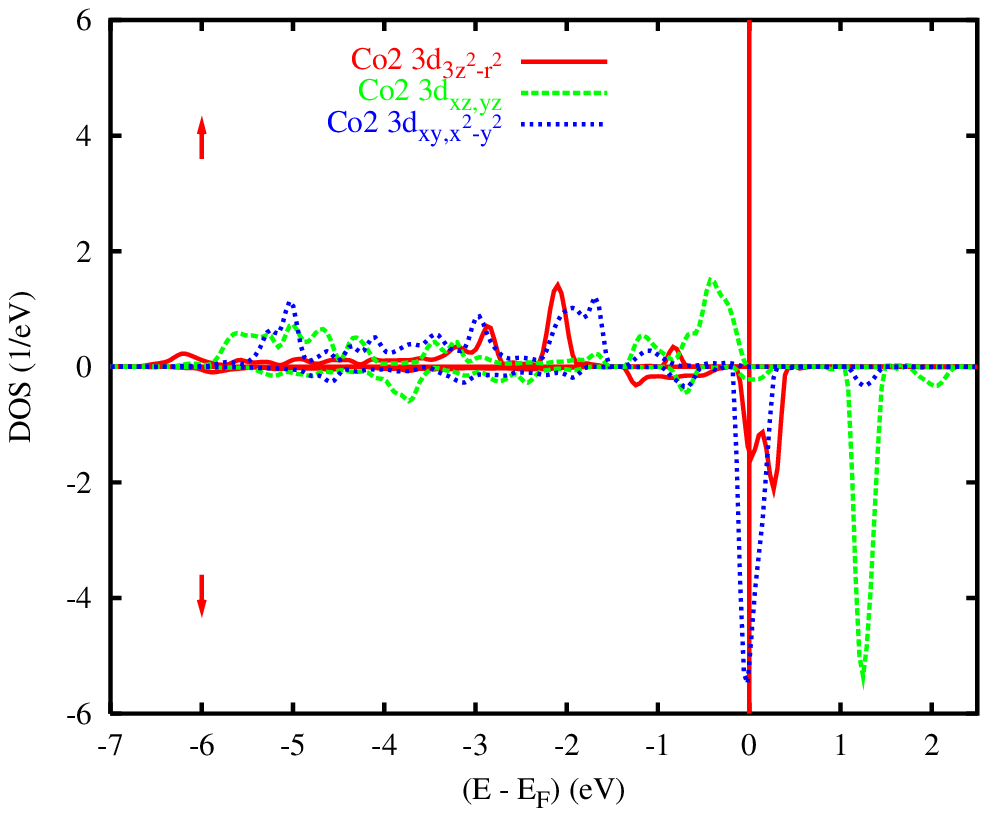}}
\caption{Partial Co1 and Co2 $ 3d $ DOS of assumed ferromagnetic 
         $ {\rm Ca_3Co_2O_6} $.}
\label{fig:res3}
\end{figure}

The Co $ 3d $ partial DOS given in Fig.\ \ref{fig:res3} 
display a rather complex relationship between the electronic and magnetic 
properties. Due to the reduced magnetic moment the Co1 $ 3d $ states 
show only small spin splitting of the order of 0.3\,eV, hence, much smaller 
than the crystal field splitting. As a consequence, the magnetic moment 
is carried alone by the $ t_{2g} $ states. Within this group of bands 
spin-majority $ d_{3z^2-r^2} $ states experience a pronounced 
localization, whereas the spin-minority states have a much larger band 
width. For the remaining $ t_{2g} $ states things are reversed. Co1 $ d $ 
contributions below -3.5\,eV display small spin-splitting but give only a 
negligible contribution to the magnetic moment. This is related to the 
fact that $ d $-$ p $ overlap within the octahedra is strongest for 
states of $ e_g $ symmetry, which are completely occupied or empty. 
The situation is different at the trigonal prismatic sites, where the 
spin-splitting is of the same order of magnitude as the crystal field 
splittings. As a consequence, spin-majority states are shifted to the 
energy region, where the oxygen $ 2p $ states dominate. At the same 
time, these states experience considerable delocalization. In contrast, 
spin-minority states form sharp peaks at and above the Fermi energy 
but give only minor contributions to the occupied states. 

\begin{figure}
\centering 
\subfigure{\includegraphics[width=0.48\textwidth]{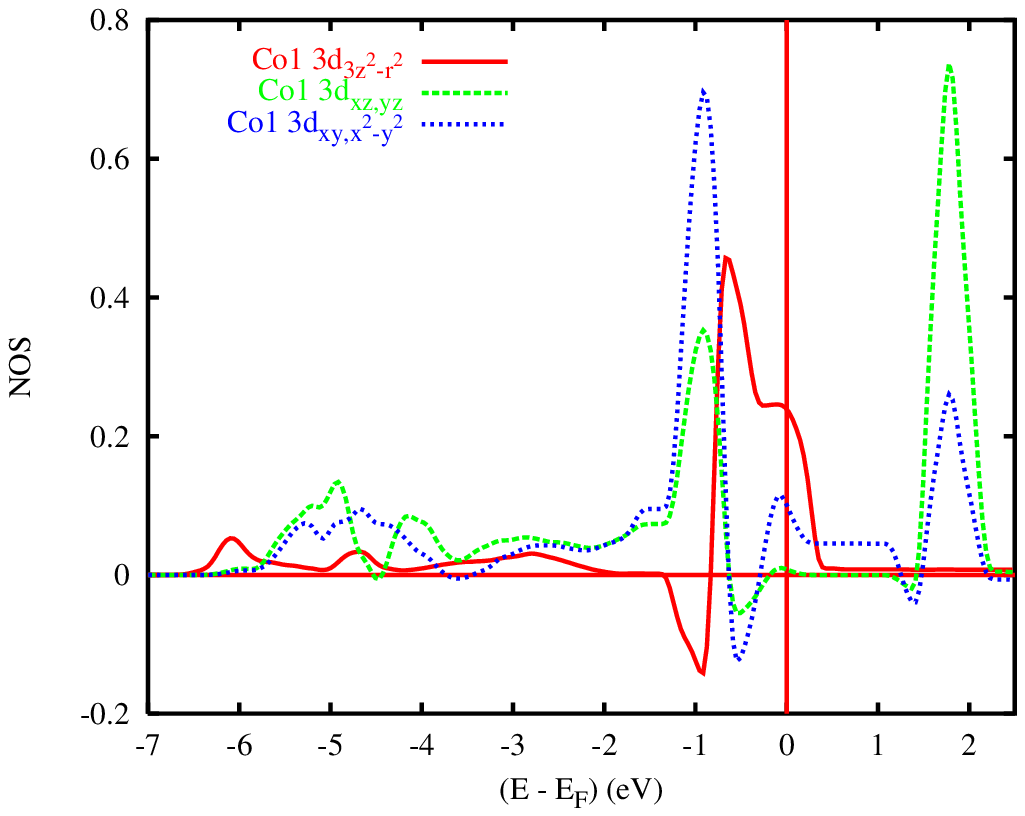}}
\subfigure{\includegraphics[width=0.48\textwidth]{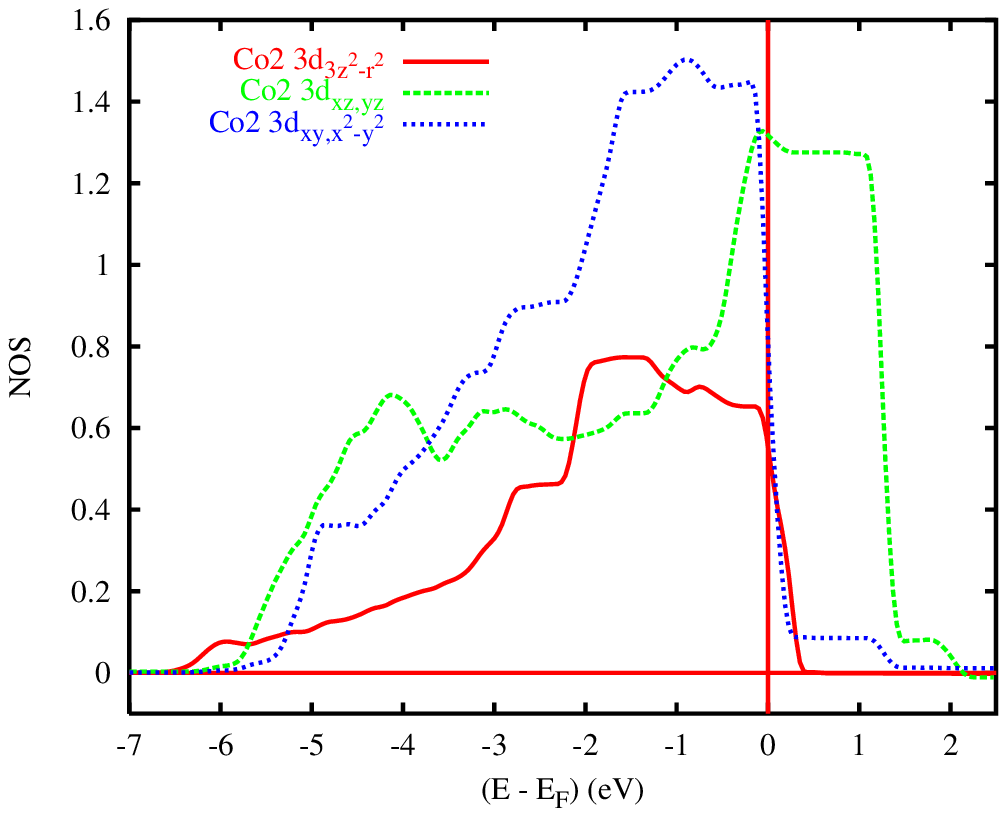}}
\caption{Partial Co1 and Co2 $ 3d $ integrated moments of assumed ferromagnetic 
         $ {\rm Ca_3Co_2O_6} $.}
\label{fig:res4}
\end{figure}

The different distributions of the electronic states in the octahedra 
and trigonal prisms have interesting consequences for the magnetic 
moments. This becomes clearer from Fig.\ \ref{fig:res4}, 
which displays energy and orbital-dependent spin magnetizations, i.e.\ the 
integrated differences of the spin-dependent partial DOS. For the Co1 sites 
it is easily seen that the small magnetic moment is mainly carried by the 
$ d_{3z^2-r^2} $ states. It starts to build up in the energy region above 
-1.5\,eV and is thus due to orbitals well localized within the Co1 atoms. 
This is completely different for the trigonal prismatic sites. There the 
magnetic moment is rather isotropic, i.e.\ it grows out of similar amounts 
from all five $ d $ states. According to Fig.\ \ref{fig:res4} the 
$ d_{3z^2-r^2} $ orbitals contribute $ 0.6 \mu_B $, while the doubly  
degenerate $ d_{xz,yz} $ and $ d_{x^2-y^2,xy} $ states give $ 1.3 \mu_B $ 
and $ 0.8 \mu_B $, respectively. Furthermore, as is clearly seen in Fig.\ 
\ref{fig:res4}, all these moments start to build up already well within 
the energy region, where the oxygen states are dominating, thus reflecting 
the considerable oxygen polarization. To conclude, the strong $ p $-$ d $ 
hybridization within the trigonal prisms, which has a strength between 
the rather nonbonding $ \pi $-type and the directed $ \sigma $-type overlap 
of the octahedra, leads to the formation of a combined isotropic 
Co2 $ d $-O $ p $ 
magnetic moment. Note that the oxygen $ p $ orbitals participate in the 
local moments and not in the exchange interaction between the prisms. 
This situation goes under the name extended moment formation and has been 
previously observed in copper oxides \cite{labacuo,weht98}. Finally, 
coupling between these extended localized moments is mediated by 
ferromagnetic exchange interaction via the $ d_{3z^2-r^2} $ orbitals of 
the low-spin Co1 atoms.

\subsection{Spin-Polarized Calculations with Ferrimagnetic Order}

In a last step, we turn to the ferrimagnetic structure, where 1/3 of the 
chains have their spins antiparallel to the remaining chains. The results 
obtained from these calculations are very similar to those for the 
hypothetical ferromagnetic structure and, hence, provide an a posteriori 
justification for the validity of the latter. Local magnetic moments of 
2.72, 0.34, and 0.14\,$ \mu_B $, respectively, are found for Co2, Co1, 
and oxygen and the total moment per unit cell again amounts to 
$ \approx 8.00 \mu_B $. According to the calculated magnetic moments as 
well as the partial DOS, which likewise look very similar to those obtained 
for the ferromagnetic situation, the differences between the spin up and 
down chains are rather small. Yet, slightly sharper peaks, i.e.\ stronger 
localization of the Co $ 3d $ states, are observed for the minority chains. 
The general situation obtained for the ferrimagnetic structure is thus 
very similar to that grown out of the calculations for ferromagnetically 
coupled chains.

\section{Conclusion}

In conclusion, electronic structure calculations for the magnetic chain 
compound $ {\rm Ca_3Co_2O_6} $ have revealed strong influence of the 
local coordination polyhedra on the electronic and magnetic properties. 
In particular, shorter bond lengths and strong $ \sigma $-type bonding 
in the octahedra allow for only small spin-splitting at these sites. In 
contrast, due to smaller crystal field splittings in trigonal prismatic 
surrounding the Co2 atoms are found in a high-spin state. Strong $ d $-$ p $ 
hybridization causes rather large oxygen magnetic moments, which, together 
with the cobalt moments at the trigonal prismatic sites, take part in the 
formation of extended but still well localized magnetic moments. Moreover, 
intrachain exchange coupling between these effective moments is based on 
metal-metal overlap of the cobalt $ d_{3z^2-r^2} $ orbitals along the 
chains and mediated by ferromagnetic exchange interaction via the 
$ d_{3z^2-r^2} $ orbitals of the low-spin cobalt atoms. Recently, mapping 
these results onto a Heisenberg model we were able to underline the 
importance of the interplay of different crystal field splittings,  
$ d $-$ p $ hybridizations, and metal-metal overlap for the ferromagnetic 
intrachain order \cite{fresard03}. Finally, according to a detailed 
analysis of the oxygen magnetic moments as well as the analysis within 
the Heisenberg model the interchain coupling seems to grow out of 
super-superexchange via oxygen states.

\section*{Acknowledgments}
We are indebted to D.\ Khomskii and A.\ Maignan for fruitful  discussions. 
C.\ Laschinger gratefully acknowledges a Marie Curie fellowship of the 
European Community program under number HPMT2000-141. This work was 
supported by the Deutsche Forschungsgemeinschaft (DFG) through 
Sonderforschungsbereich SFB 484 and by the BMBF (13N6918A).

\end{document}